\documentstyle[prd,aps,eqsecnum,amsfonts,graphicx]{revtex}

\begin{document}

\twocolumn[\hsize\textwidth\columnwidth\hsize\csname
@twocolumnfalse\endcsname
\date{December 7, 1999}
\rightline{TPI-MINN-99/58-T}
\rightline{UMN-TH-1831/99}
\rightline{FTUV/99-81}
\rightline{IFIC/99-85}

\title{Domain wall junctions in a generalized
Wess-Zumino model.}
\author{Daniele Binosi}
\address{
Departamento de F\`\i sica Te\'orica and IFIC, Centro Mixto, 
Universidad de Valencia-CSIC,\\
E-46100, Burjassot, Valencia, Spain\\
$\mathrm e$-$\mathrm mail: binosi@titan.ific.uv.es$}
\author{Tonnis ter Veldhuis}
\address{
Theoretical Physics Institute,
School of Physics and Astronomy,
University of Minnesota, 
116 Church St,\\
Minneapolis, MN 55455,
U.S.A. \\
$\mathrm e$-$\mathrm mail: veldhuis@hep.umn.edu$}
\maketitle
\begin{abstract}
We investigate domain wall junctions in a generalized Wess-Zumino
model with a ${\mathbb Z}_N$ symmetry. We present a method to identify the junctions
that are potentially BPS saturated. We then use a numerical
simulation to show that those  junctions indeed saturate the
BPS bound for $N=4$.
In addition, we study the decay of unstable non-BPS junctions.
\end{abstract}
\vskip2pc]

\section{Introduction.}

In field theories with discrete, degenerate vacua,
domain walls -- field configurations
that interpolate between different vacua -- may occur.
Supersymmetric theories,
for which the vacuum energy is positive semi-definite, frequently
have degenerate vacua. In many instances, multiple
vacua exist in which the vacuum energy vanishes and supersymmetry
remains unbroken. The degenerate vacua may or may not be related by a 
spontaneously broken discrete symmetry.

General considerations concerning domain walls in supersymmetric
theories are presented in \cite{DS2,CS,AT}. 
A lower bound on the tension, the BPS bound, can be calculated
without explicitly solving for the profile of the wall.
Domain walls break
the translation symmetry in one direction, and
supersymmetry is either completely broken, or $1/2$ of it
is preserved. 
In the latter case, a central extension appears in the
supersymmetry algebra. The tension saturates the BPS bound and
is equal to the $(1,0)$ central charge.

In \cite{GT,CHT} it was noted that more complicated field configurations
with axial geometry, domain wall junctions, may also occur. 
These configurations, of the ``hub and spoke'' type, are a natural 
generalization of domain walls, and 
interpolate between more than two degenerate vacua. Far away from the center
of the junction, the fields are approximately constant -- at the 
vacuum expectation values of the various vacua -- 
in sectors. These sectors
are separated by ``spokes'', where the field interpolates
between one vacuum to the next following  approximate domain wall
profiles. In the center of the junction, ``the hub'', the domain
walls meet and the field configuration resembles a string. If these 
junctions are BPS saturated, $1/4$ of the original
supersymmetry is preserved and both the $(1,0)$ and $(1/2,1/2)$ central
charges  appear in the supersymmetry
algebra. The junction tension is determined by
a combination of the $(1/2,1/2)$ and $(1,0)$ central charges \cite{GS}. 
Although the two central charges contain (related) ambiguities, it is 
shown in \cite{GS,SV}
that the ambiguities cancel in physical quantities like the tension.

It is worthwhile to emphasize that domain walls in supersymmetric models 
are not necessarily BPS saturated.
For example, in the model of \cite{Sm} the BPS saturation
of a domain wall depends on the values of the model parameters. 
In many models it is not possible to obtain explicit analytical solutions
to the first order BPS equations.  In those models the question
whether solutions exist at all can still be answered. In \cite{FMVW}
a necessary and sufficient condition which indicates whether 
two vacua are connected by a PBS saturated domain wall was 
developed\footnote{It is not
obvious that this method works also for theories with more than one chiral
superfield.}.  

The issue of BPS saturation in junctions is  more complicated,
because a similar condition does not exist.
At this time, there is only one known model \cite{OINS} in which a domain wall
junction was found as an explicit analytical solution to the
BPS equations. In the absence of
analytical expression for the field configuration, alternative
methods must be employed to study junctions.
Some information can be obtained
from analytical results for specific limiting
values of parameters or coordinates. Complementary results
can be obtained from numerical simulation
of the equations of motion.

For any domain wall junction, a BPS bound on the junction tension
can be calculated. In \cite{SV} the general method to obtain
this bound from the domain walls surrounding the junction is
presented. In this work, it was also established that the junction 
tension is negative in general, although ``exotic'' models  in which
domain wall junctions have 
positive tension may exist.

In this paper we will study domain wall junctions in a class of 
generalized Wess-Zumino (WZ) models with ${\mathbb Z}_N$ symmetry.
The Lagrangian is given by
\begin{equation}
{\mathcal L}=\int\!d^2\theta\int
\!d^2\bar\theta\,{\mathcal K}(\Phi,\Phi^\dagger)
+\left\{\int\!d^2\theta\,{\mathcal W}(\Phi)+h.c.\right\},
\end{equation}
where the K\"ahler potential has the canonical form 
${\mathcal K}(\Phi,\Phi^\dagger)=\Phi
\Phi^{\dagger}$, and the superpotential is given by\footnote{
More generally we could consider the superpotential
${\mathcal W}(\Phi)=A \Phi - B \Phi^{N+1}/(N+1)$, but for our purposes this is 
equivalent to Eq.~(\ref{superpot}) by rescaling the field
and the coordinates as
$\phi\rightarrow (A/B)^{1/N} \phi$ and $x_{\mu}\rightarrow
A^{(1-N)/N} B^{-1/N} x_{\mu}$.
In the literature, the large $N$ limit of the model is studied
with $A=N$ and $B=1/N^{N-1}$.} 
\begin{equation}
{\mathcal W}\left(\Phi\right)=\Phi-\frac{1}{N+1}\Phi^{N+1}. \label{superpot}
\end{equation}
The action is invariant under the transformation 
$\Phi(x,\theta)\rightarrow {\mathrm e}^{2 \pi\imath/N}
\Phi(x,\theta {\mathrm e}^{-\pi\imath/N})$. Moreover,
the model has $N$ physically equivalent vacua
given by
\begin{equation}
\phi={\mathrm e}^{2 \pi\imath k/N},\qquad k=0,1,\dots,N-1,
\label{vacua}
\end{equation}
where $\phi$ is the scalar component of the superfield $\Phi$.
Despite its apparent simplicity, the model is of physical interest
because it is related to the Veneziano-Yankielowicz action \cite{VY}, 
which is a low energy effective action for 
supersymmetric QCD. The parameter $N$ is related to the number
of colors in the ${\rm SU}(N)$ gauge group. 

For $N=2$, the model reduces to the renormalizable Wess-Zumino model.
It is well known that there is an analytical expression for the domain wall 
in this case (see \cite{CS}, for example). In the large $N$ limit,
an analytical expression for the ``basic'' domain wall --
the wall that connects vacua for consecutive values of $k$ --
is presented in \cite{DGK,DK} to next to leading 
order in $1/N$. In this reference it is also noted that all
vacua in  the model are connected by BPS saturated domain walls at finite $N$.

The ``basic'' domain wall junction has also been studied.
We define the ``basic'' junction as 
the field configuration that cycles through all $N$ vacua,
consecutively from $k=0$ to $k=N-1$, counter-clockwise around
the center of the junction.
In \cite{GaS} an analytical solution to the BPS equation
for the ``basic'' junction in the $N\rightarrow \infty$ limit
was presented. In \cite{SV} it was
confirmed that the ``basic'' junction is also BPS saturated for
finite values of $N$ by numerical simulations. The tension of the 
``basic'' junction was also calculated, analytically to next
to leading order in $1/N$, and numerically
also for finite values of $N$. 

Beside the ``basic'' junction, the vacua of the model allow
for a multitude of other types of junctions.
As discussed further in the paper, we devise a general method to
identify all potential BPS saturated junction in this model. 
There is a very stringent
(asymptotic) consistency condition which allows only for a well defined set of
intersection angles between the walls. The set of potential
BPS junctions we identify contains the junctions that appear
in \cite{Sa,BB}, where tilings of domains and networks
of domain walls in this model are studied.

We consider four types of domain wall junctions in the model
with $N=4$, two of which belong to the class of potential
BPS saturated, and two more that are not  BPS saturated.
For each junction we perform a numerical simulaton of the
second order equations of motion to obtain the junction profile. 
For the
potential BPS junctions, we compare the energy of the junction 
to the BPS bound (which we evaluated from the
domain walls surrounding the junction by the methods in \cite{SV}) to determine
whether they are indeed BPS saturated. For the other two
junctions, we investigate their stability and decay.
Although we perform our numerical calculations for the 
value $N=4$, we expect our results to apply to other
values as well. Moreover, the methods we use are
quite general, and can be applied to other models. 

The remainder of the paper is organized as follows.
In Sec.~\ref{sec:dw}, we describe the possible domain walls of
the model. 
Then in Sec.~\ref{sec:tIdw} and~\ref{sec:tIIdw}, we describe in
detail the two types of domain walls which exist in the case $N=4$.
These walls 
appear far a away from the center of the junctions that
we consider. 
In Sec.~\ref{sec:j} we
first discuss the potential BPS junctions in the models for  generic $N$,
which satisfy a very restrictive consistency condition. We then focus
on the $N=4$ case and describe in detail the two types of
BPS junctions (Sec.~\ref{sec:tAj} and~\ref{sec:tBj}) and the non-BPS types
(Sec.~\ref{sec:nbj}) which occur. In Sec.~\ref{sec:nm} we give a brief outline of the
computational scheme used for obtaining the numerical representation of the 
junctions; the
paper ends with some comments in Sec.~\ref{sec:s}.

\subsection{Domain walls.}\label{sec:dw}
BPS saturated domain walls are solutions to the equation
\begin{equation}
\partial_x\phi = {\mathrm e}^{\imath \delta} (1-\bar{\phi}^N),
\label{bpswall}
\end{equation}
with the boundary condition that $\phi$ approaches vacuum
values at $x \rightarrow -\infty$ and $x \rightarrow \infty$.
Such solutions exist for particular values of the
phase $\delta$, which are determined by the equation
\begin{equation}
{\mathrm Im}\left\{{\mathrm e}^{-\imath \delta} {\mathcal W}(\phi(x\rightarrow -\infty))
\right\}={\mathrm Im}\left\{{\mathrm e}^{-\imath \delta} {\mathcal W}(\phi(x\rightarrow 
+\infty))\right\}. 
\label{deltaeq}
\end{equation}

For the model with $N=4$, there are two types of domain walls.
The first type connects adjacent vacua, for which $k$ differs
by one. The second type connects opposite vacua, for which
$k$ differs by two. We will discuss each of these types in turn.

\subsubsection{Type~I domain walls.}\label{sec:tIdw}
Although it is not possible to find an analytical solution
for the domain wall profile, the characteristic features can be
readily uncovered. For definiteness, we focus on the wall that
connects the vacua
labeled by $k=0$ and $k=1$.  From Eq.~(\ref{deltaeq}) it follows
that for this wall the value of $\delta$ is either 
$-\pi/4$ or $3 \pi/4$. 
For $\delta=-\pi/4$ ($3\pi/4$) the wall connects
the vacuum with $k=1$ ($k=0$) at $x=-\infty$ to the vacuum with $k=0$
($k=1$) at $x=\infty$. 

A general solution to the BPS equation
contains one integration constant, indicating the freedom
to shift the center of the wall. We fix this constant
by choosing the center of the wall to be at the origin: The
wall is then invariant under the transformation 
$\phi(x)=\imath \bar{\phi}(-x)$. To study the wall close to the
center, it is useful to write the field in terms of polar
variables, $\phi = \rho\,{\mathrm e}^{\imath \alpha}$. The
symmetry of the wall dictates that $\alpha=\pi/4$ and 
$\partial_x\rho=0$ at the center. It then follows from the BPS equation
that, for $\delta=-\pi/4$,
\begin{equation}
\left. \partial_x\alpha\right|_{x=0} = - \frac{1+\rho_0^4}{\rho_0},
\end{equation} 
where $\rho_0=\rho(0)$. The value of $\rho_0$ can be calculated
by evaluating the constant of motion along the wall, {\it i.e.} ${\mathrm Im}\left\{
{\mathrm e}^{-\imath
\delta} {\mathcal W}(\phi)\right\}$ at $x=\infty$ and at $x=0$. As a consequence,
$\rho_0$ is the sole real root of the equation
\begin{equation}
\rho_0 + \frac{\rho_0^5}{5} = \frac25 \sqrt{2}.
\end{equation}
Numerically, $\rho_0=0.55514$. The tension
of the wall can be calculated exactly, and depends only on
the value of $\phi$ at the two ends of the wall; it is given by
\begin{equation}
T_1^{(I)}=2 \left| {\mathcal W}(\phi(x \rightarrow \infty)) - 
{\mathcal W}(\phi(x\rightarrow -\infty)) \right| = \frac85 \sqrt{2}.
\end{equation}
In Table~I we present all 
domain walls of type~I with their corresponding values for the phase
$\delta$.\\

\begin{center}
\begin{tabular}{|c|c|c|}
\hline \hline \hline
$\delta$ & $\hspace{2.6cm} \phi(-\infty)$ & $\hspace{2.6cm} \phi(\infty)$\\
\hline \hline
$-3\pi/4$ & \hspace{2.6cm} $1$       & \hspace{2.6cm} $-\imath$\\
          & \hspace{2.6cm} $\imath$  & \hspace{2.6cm} $-1$\\
$-\pi/4$  & \hspace{2.6cm} $\imath$  & \hspace{2.6cm} $1$\\
          & \hspace{2.6cm} $-1$      & \hspace{2.6cm} $-\imath$\\
$\pi/4$   & \hspace{2.6cm} $-1$      & \hspace{2.6cm} $\imath$\\
          & \hspace{2.6cm} $-\imath$ & \hspace{2.6cm} $1$\\
$3\pi/4$  & \hspace{2.6cm} $-\imath$ & \hspace{2.6cm} $-1$\\
          & \hspace{2.6cm} $1$       & \hspace{2.6cm} $\imath$\\
\hline \hline \hline
\end{tabular}\\
\end{center}
{\small{\hspace{4pt} TABLE I. Type~I walls with their corresponding values 
of~$\delta$.}}

\subsubsection{Type~II domain walls.}\label{sec:tIIdw}
For definiteness, we focus on the wall connecting the vacua labeled 
by $k=2$ and $k=0$. For this wall, the value of $\delta$ is either
$0$ or $\pi$. For $\delta=0$ ($\delta=\pi$) the wall connects
the vacuum with $k=2$ ($k=0$) at $x\rightarrow-\infty$ to the vacuum
with $k=0$ ($k=2$) at $x\rightarrow \infty$. The wall
is purely real. Again, we fix the integration constant
so that the wall is centered at the origin; with this choice, 
the wall is invariant under the transformation $\phi(x)=-\phi(-x)$. 

The field vanishes at the origin, its derivative being (for $\delta=0$)
\begin{equation}
\left. \partial_x\phi\right|_{x=0}=1,
\end{equation}
while the wall's tension in this case is
\begin{equation}
T_1^{(I)}=2 \left| {\mathcal W}(\phi(x \rightarrow \infty))- 
{\mathcal W}(\phi(x\rightarrow -\infty)) \right| = \frac{16}5. 
\end{equation}
Table~II lists all 
domain walls of this type with their corresponding value of
$\delta$.

\begin{center}
\begin{tabular}{|c|c|c|}
\hline \hline \hline
$\delta$ & $\hspace{2.8cm} \phi(-\infty)$ & $\hspace{2.7cm} \phi(\infty)$\\
\hline \hline
$0$ 		& \hspace{2.8cm} $1$       & \hspace{2.7cm} $-\imath$\\
$\pi/2$  	& \hspace{2.8cm} $\imath$  & \hspace{2.7cm} $1$\\
$\pi$   	& \hspace{2.8cm} $-1$      & \hspace{2.7cm} $\imath$\\
$3\pi/2$  	& \hspace{2.8cm} $-\imath$ & \hspace{2.7cm} $-1$\\
\hline \hline \hline
\end{tabular}\\
\end{center}
{\small{\hspace{4pt} TABLE II. Type~II walls with their corresponding values 
of~$\delta$.}}

\subsection{Junctions.}\label{sec:j}
BPS junctions are solutions to the equation
\begin{equation}
\partial_z\phi= \frac{{\mathrm e}^{\imath \delta}}2 (1-\bar{\phi}^N).
\label{difj}
\end{equation}
Here $z=x+\imath y$ and $\partial_z=1/2
(\partial_x - \imath \partial_y)$.
Junctions are static configurations that have the geometry
of a hub and
spokes system. 
In the sectors between the spokes, the field $\phi$ approaches
the vacuum values; far away from the hub, $\phi$ follows domain
wall profiles from one sector to the next on trajectories
perpendicular to the spokes.
Each spoke is therefore associated with a domain wall. The
hub is the center of the junction where the domain walls meet.

A few considerations are important to consider what junctions
are possible, and which of those junctions are BPS saturated. 

First, the sum of the forces exerted on the junction by the domain walls
that are attached to it has to vanish for a static configuration. 

Second, the value of the phase $\delta$ in the BPS equation for the
junction merely determines
the orientation of the junction. The effective value of 
the phase for the BPS domain wall equation for 
a wall associated with a spoke that makes an 
angle $\alpha$ with the $\hat{y}$ axis
is equal to $\delta-\alpha$. Once the direction of 
one spoke is fixed, for example to point in the $\hat{y}$ direction, 
the value of $\delta$ is fixed. For the remainder of the
spokes, only specific values of $\alpha$ are then possible. 
This restriction on the possible values of $\alpha$ turns out
to be a very stringent consistency condition.
In the case of the
model with $N=4$, inspection of Tables~I and~II 
shows that this condition allows only two types of 
junctions as potential BPS candidates. We call these two types
A and B respectively, and we investigate them below.

Once it has been established that a static junction exists, another
interesting issue to determine is whether or not it is stable. If a
junction is BPS saturated, it is guaranteed to be stable (also see the
discussion in \cite{SV}). For
non-BPS junctions, the issue is not so clear; if the boundary
conditions also allow a BPS saturated configuration, then the
non-BPS junction is at most meta-stable.

One can actually devise a general method for generating all the potential BPS
saturated junctions in our class of models.
In fact, given the presence of the $N$ degenerate vacua (\ref{vacua}),
one has that for $N$ even (odd) there are $N/2$ ($(N-1)/2$) 
types of domain walls.
These different types of walls can be labeled by the integer
$l=|k_1-k_2|$ or $l=N-|k_1-k_2|$, whichever is smaller (here
$k_1$ and $k_2$ indicate the vacua that are connected by
the wall); moreover the tension of a wall,
labeled by $l$, is equal to
\begin{equation}
T^{(l)}= \frac{2N}{N+1} \sqrt{2 - 2 \cos{\frac{2\pi l}N}}.
\end{equation} 

Now, we can represent each type of wall by a vector with length
equal to its tension; thus a junction will be represented by
the vectors of the walls that it contains far away from
the origin, with the direction of the vectors chosen so that
they point in the direction where the corresponding wall is located.
In a sense, the vectors play the role of the spokes in the
hub and spoke system, but they have additional information 
because of their length. Crossing a spoke counter clockwise means
that the value of $k$ increases by $l$ from one domain to the
next along the corresponding domain wall: Thus, two junctions will be equivalent
if they can be mapped onto each other by inversions and rotations.

\begin{figure}[!h]
\begin{center}
\includegraphics[width=4cm]{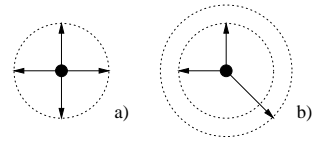}
\end{center}
\caption{Potential BPS saturated junctions for the case $N=4$.}
\label{N4}
\end{figure}

\begin{figure}[!h]
\begin{center}
\includegraphics[width=8cm,angle=-90]{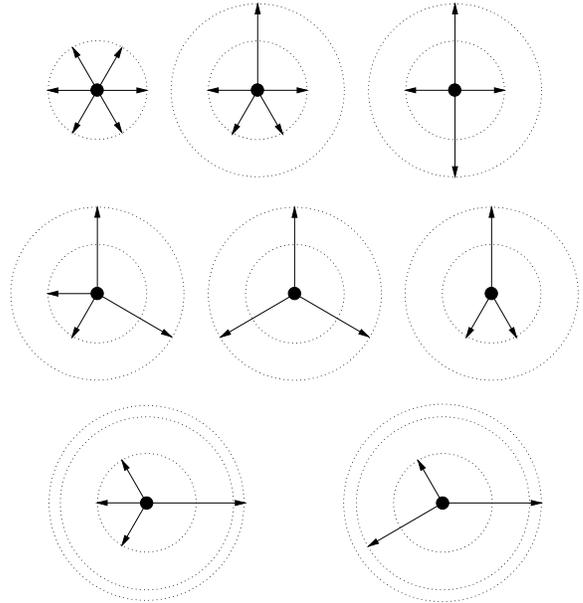}
\end{center}
\caption{Potential BPS saturated junctions for the case $N=6$.}
\label{N6}
\end{figure}

All of the potential BPS saturated junctions can then be obtained as follows.
We start with the basic $N$-junction, which consists of $N$ vectors 
representing walls with $l=1$ (type~I walls) at angles of $2\pi/N$. Then we add any two
neighboring vectors in the diagram representing the junction, repeating
this process until all unequivalent diagrams have been 
generated. For $N=4$ this procedure gives the type~A and B junctions of above 
(Fig.~\ref{N4}). In Fig.~\ref{N6} all the potential
BPS junctions are shown for the case $N=6$.

\subsubsection{Type~A junction.}\label{sec:tAj}
Here we study the junction of the type shown in Fig.~\ref{N4}~a).
Far away from the center of  the junction, there are four sectors in
which the field consecutively takes on the four possible vacuum 
expectation values.
These sectors are separated by BPS domain walls of type~I, which 
were discussed before.

Near the center of the junction, the field can be expanded in a
power series in $z$ and $\bar{z}$.
Accordingly one has
\begin{equation}
\phi(z,\bar z)=\sum_{n=0}^\infty\sum_{m=0}^\infty a_{mn}z^n\bar z^m,
\label{exp}
\end{equation}
with $a_{mn}$ some complex coefficients\footnote{One can in principle allow also
for negative $n$ and $m$ with the condition $m>\vert n\vert$ ($n>\vert m\vert$) 
if $n<0$ ($m<0$). These negative values are however excluded when one substitute
the expansion (\ref{exp}) in the differential equation (\ref{difj}).}.

The differential equation (\ref{difj}) and the boundary
conditions are invariant under the transformations ($\gamma=-\pi/2+\pi/N$)
\begin{eqnarray}
{\mathbb Z}_2 & : & \ \phi(z,\bar z) \rightarrow 
\phi^{\dagger}({\mathrm e}^{-2\imath\gamma} \bar z,{\mathrm e}^{2\imath\gamma}z); \\
{\mathbb Z}_N & : & \ \phi(z,\bar z) \rightarrow 
-{\mathrm e}^{-2\imath\gamma} \phi(-{\mathrm e}^{2\imath\gamma} z,-{\mathrm
e}^{-2\imath\gamma}
\bar z).
\end{eqnarray}

Imposing the ${\mathbb Z}_N$ symmetry the expansion  (\ref{exp}) takes the 
form
\begin{equation}
\phi(z,\bar z)=\sum_{k=0}^\infty\sum_{l=0}^\infty\left[b_{kl}z^{1+Nk}(z\bar
z)^l+c_{kl}\bar z^{N(k+1)-1}(z\bar z)^l\right],
\end{equation}
with $b_{kl}$ and $c_{kl}$ again complex coefficients. Imposing in
addition the ${\mathbb Z}_2$ symmetry, we find the following constraints on
these coefficients
\begin{eqnarray}
b_{kl} & = & (-1)^{Nk}{\mathrm e}^{2\imath\gamma}b^\dagger_{kl}, \\
c_{kl} & = & (-1)^{N(k+1)}{\mathrm e}^{2\imath\gamma}c^\dagger_{kl}.
\end{eqnarray}

Thus, if $N$ is even, one can introduce the real coefficients $d_{kl}$ and
$e_{kl}$, obtaining the expansion
\begin{equation}
\phi(z,\bar z)={\mathrm e}^{\imath\gamma}
\sum_{k=0}^\infty\sum_{l=0}^\infty\left[d_{kl}z^{1+Nk}
(z\bar z)^l+e_{kl}\bar z^{N(k+1)-1}(z\bar z)^l\right]. 
\end{equation} 
If on the other hand $N$ is odd, one has
\begin{eqnarray}
b_{kl} & = & (-1)^k{\mathrm e}^{2\imath\gamma}b^\dagger_{kl}, \\
c_{kl} & = & (-1)^{(k+1)}{\mathrm e}^{2\imath\gamma}c^\dagger_{kl},
\end{eqnarray}
so that, according to the value of the index
$k$, half of the coefficients is real, the other half being purely
imaginary.
In the case $N=4$ one has $\gamma=-\pi/4$, so that
\begin{equation}
\phi(z,\bar z)={\mathrm e}^{- i \pi/4 }
\sum_{k=0}^{\infty}\sum_{l=0}^{\infty}\left[
d_{kl} z^{1+4k} (z\bar z)^l +
e_{kl} {\bar z}^{3+4k} (z\bar z)^l\right].
\label{req}
\end{equation}

We can now calculate the coefficients $d_{kl}$  and $e_{kl}$ using the 
BPS equation; it turns out that all coefficients are
determined except for the ones which multiply terms that
contain only $\bar z$, that is, the $e_{k0}$. These coefficients have
to be adjusted to obtain the correct domain walls far from the center
of the junction\footnote{Note that the expansion up to terms
of eleventh order in $z$ and $\bar z$ contains only two adjustable 
parameters.}.

\begin{figure}[!h]
\begin{center}
\includegraphics[width=7cm]{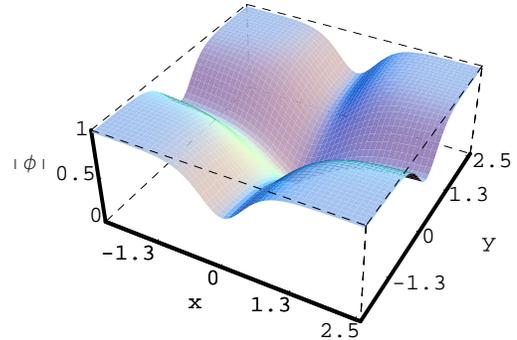}
\end{center}
\caption{Modulus of $\phi$ for the type~A junction.}
\label{4j}
\end{figure}

Close to the origin, the configuration looks like a string.
We have proved then that there are sensible solutions to the BPS equations
both near the origin and far away from it. In order
to determine whether those asymptotic solutions can be
connected consistently and thus to investigate whether the
junction is BPS saturated we performed a numerical simulation
of the second order equations of motions for the field
$\phi$, while imposing the domain walls as the boundary
conditions at large distance from the origin. In order
to allow the field to relax to its minimum energy configuration,
we also introduced a damping term (see Sec.~\ref{sec:nm} for details).
In Fig.~\ref{4j} we show the modulus of $\phi$ after the configuration
has come to rest. 

In order to determine whether the
junction is BPS saturated, we will first calculate the
BPS bound on the energy inside a square of ``radius'' $R$, with
the center of the junction located at the center of the 
square. Assuming the junction to be BPS saturated
the energy inside
a square with radius $R$ can be calculated entirely in terms of 
quantities related to the domain walls surrounding the junction
when the radius $R$ is much larger than the size of the junction.
Explicitly, for large $R$, the bound on the energy takes the form
\begin{equation}
\frac{E_{{\mathrm BPS}}}L = T_2^{(A)} + 4 T_1^{(I)} R.
\label{bound}
\end{equation}

Here $L$ is a distance in the direction perpendicular to
the $x$-$y$ plane. The second term in (\ref{bound}) is the contribution to
the energy from the four type~I domain walls connected
to the junction; the first term represents instead the energy of the junction,
which is equal to
\begin{equation}
T_2^{(A)} = \oint_{{\mathrm square}}^{R\gg 1} \vec{a} \cdot d\vec{l},
\end{equation}
where
\begin{equation}
a_i= {\rm{Im}} \left\{ \phi \partial_{x_i}\bar\phi \right\}.
\end{equation}

Now, each wall on the contour of the square contributes
the same amount to the tension. The BPS bound on the junction tension is 
therefore equal to
\begin{equation}
T_2^{(A)} = 4 \int_{-\infty}^{\infty} a^{(I)} dx,
\end{equation}
where $a^{(I)}= {\rm{Im}} \left\{ \phi^{(I)} \partial_x\bar\phi^{(I)} 
\right\}$
and $\phi^{(I)}$ is the profile of a type~I wall along
the $\hat{x}$ axis.
Using a numerical representation of the type~I domain wall, we determined that
$T_2^{(A)}= -2.8614$. 

\begin{figure}[!h]
\begin{center}
\includegraphics[width=7cm]{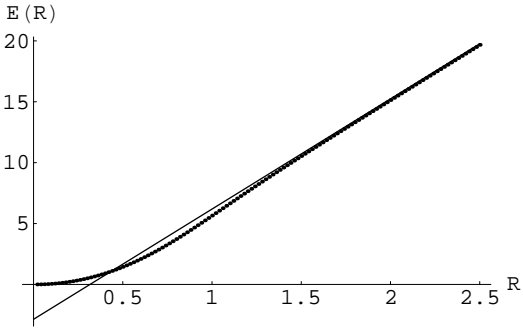}
\end{center}
\caption{Energy inside a square of radius $R$ as a function of $R$ (dots)   
compared to the BPS bound (solid line) for the type~A junction.}
\label{EofR1}
\end{figure}

We also calculated the energy 
in a square with radius $R$ as a function of $R$ 
directly from our numerical representation of the junction
using the equation
\begin{equation}
\frac{E}{L}= \int_{\rm{square}}^{R} \left\{ \partial_x\bar\phi \partial_x\phi
+ \partial_y\bar\phi \partial_y\phi + 
V  \right\} dx dy, \label{numen}
\end{equation}
where $V=\partial_\phi{\mathcal W} \partial_{\bar\phi}\bar{\mathcal W}$ 
is the scalar potential.
In Fig.~\ref{EofR1} we compare this energy 
with the BPS bound. It is clear that the energy converges to
the bound for large $R$ (compared to the size of the junction)
and therefore the junction does
indeed saturate the BPS bound.

\subsubsection{Type~B junction.}\label{sec:tBj}
The second type of junction that is potentially 
BPS saturated has two spokes corresponding to 
type~I walls  at an angle of
$\pi/2$. A third spoke associated with a type~II domain wall
intersects each of the other spokes at an angle of $3 \pi/4$. 
The configuration is sketched in 
Fig.~\ref{N4}~b). 

\begin{figure}[!h]
\begin{center}
\includegraphics[width=7cm]{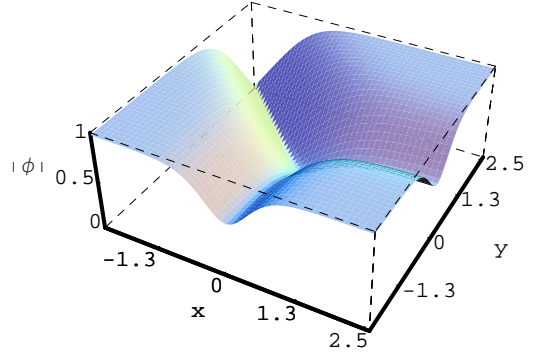}
\end{center}
\caption{Modulus of $\phi$ for the type~B junction.}
\label{3j}
\end{figure}

This type of junction may be obtained from
a junction of type~A by pinching two neighboring spokes together in
a symmetric fashion, thereby eliminating one domain. We 
performed a numerical simulation of the second order equations
of motion, with the appropriate domain wall profiles as boundary
conditions, and including a damping term (see again Sec.~\ref{sec:nm} 
for details). In Fig.~\ref{3j} we
show the modulus of the field $\phi$ after the configuration
has come to rest.

To analyze whether this type of junction is BPS saturated we
follow a procedure similar to the one used for the type~A junction.
We first calculate the bound on the energy in a square of
radius $R$ for large $R$, assuming the junction to be BPS saturated.
This bound takes the form
\begin{equation}
\frac{E_{{\mathrm BPS}}}{L} =T_2^{(B)} + 2 T_1^{(I)} R + \sqrt{2} T_1^{(II)} R
- \Delta T.
\end{equation}

The first term represents the BPS bound on the tension of the
junction. The second and the third term represent the contribution
to the energy from the two type~I walls and the type~II wall (the
factor of $\sqrt{2}$ appears because the spoke representing the
type~II wall is in the diagonal direction). Special care must be
taken because the spoke representing the type~II wall intersects 
the sides of the square diagonally, so that
the energy contributed by two triangular areas needs to be
subtracted. This is the origin of the fourth term, $\Delta T$, which
takes the form
\begin{equation}
\Delta T = 4 \int_0^{\infty} \partial_x\phi^{(II)}
\partial_x\bar{\phi}^{(II)}x dx,
\end{equation}
where $\phi^{(II)}$ is the profile of a type~II wall along the
$\hat{x}$ axis.

The field $\phi$ on the type~II wall connecting the $k=0$ sector 
to the $k=2$ sector of this junction is purely real, so that it 
does not contribute to the BPS bound on the tension of 
the junction. On the other hand the tension receives equal contributions from each
of the type~I walls. The BPS bound on the tension of type~B junctions
is therefore half of the tension of a type~A junction
\begin{equation}
T_2^{(B)}=\frac{1}{2} T_2^{(A)}.
\end{equation}

We can finally evaluate $\Delta T$ using the numerical representation of the
type~II wall, finding in this way
\begin{equation}
\Delta T= - \frac{1}{2} T_2^{(A)}.
\end{equation}
This is a curious equality, as $\Delta T$ is a quantity calculated
from a type~II wall, and $T_2^{(A)}$ can be calculated from
a type~I wall profile. We have no further comments on the
nature of the equality here, but as a consequence the total energy
in a square with large radius is equal for type~A and type~B junctions.

\begin{figure}[!h]
\begin{center}
\includegraphics[width=7cm]{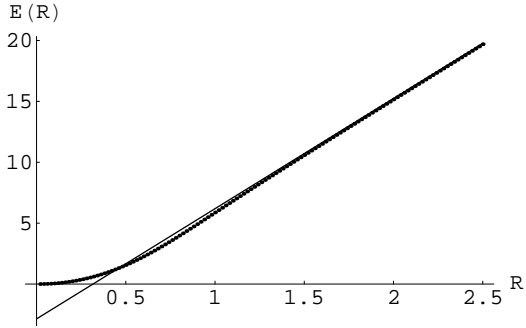}
\end{center}
\caption{Energy inside a square of radius $R$ as a function of $R$ (dots)   
compared to the BPS bound (solid line) for the type~B junction.}
\label{EofR2}
\end{figure}

We also calculated the energy 
in a square with radius $R$ as a function of $R$ 
directly from our numerical representation of the type~B junction
using Eq.~(\ref{numen}).
In Fig.~\ref{EofR2} we compare this energy 
with the BPS bound. It is clear that for this junction too,
the energy in a square converges to
the bound for large $R$ (compared to the size of the junction).
The type~B junctions therefore also
saturate the BPS bound.

\subsubsection{Non-BPS junctions.}\label{sec:nbj}

Here we discuss static junctions that are not BPS saturated.
For such junctions the values of $\delta$ associated
with the various spokes are not consistent. However, the total force
on the junction vanishes. Moreover,  they are invariant under  an
extended symmetry group, and therefore extremize the energy.

\begin{figure}[!h]
\begin{center}
\includegraphics[width=4cm]{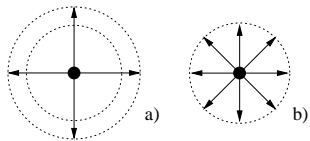}
\end{center}
\caption{Non-BPS saturated junctions in the case $N=4$.}
\label{N4notBPS}
\end{figure}

The first junction of this type that we discuss is displayed 
in Fig.~\ref{N4notBPS}~a). It consists out of
four sectors separated by spokes which intersect at $\pi/2$
angles. However, in contrast to the BPS four-junction, the
spokes are associated with type~II walls. 

\begin{figure}[!h]
\begin{center}
\includegraphics[width=7cm]{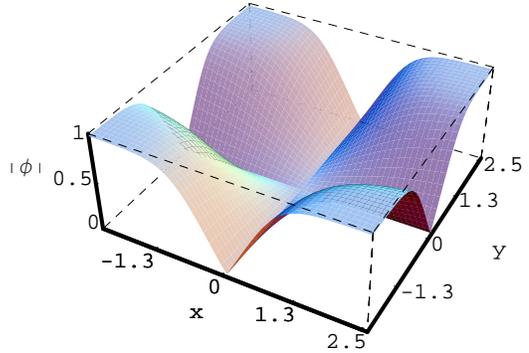}
\end{center}
\caption{Modulus of $\phi$ for the non-BPS four-junction.}
\label{sj}
\end{figure}

We found by a
numerical simulation of the second order equations of motion
that a static configuration of this type does indeed exist.
In Fig.~\ref{sj} we show the modulus of $\phi$ for the final
configuration at rest. 

\begin{figure}[!h]
\begin{center}
\includegraphics[width=7cm]{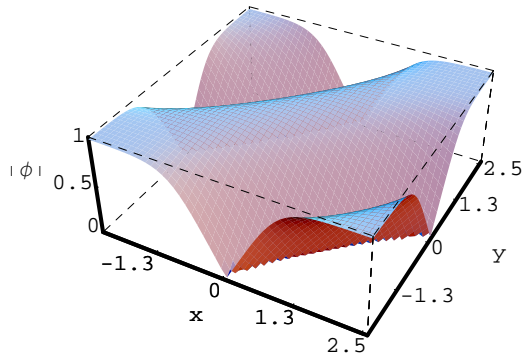}
\end{center}
\caption{Modulus of $\phi$ after the non-BPS four-junction has decayed into a
configuration with three domains separated by two walls.}
\label{asj}
\end{figure}

However, this junction has a negative
mode. It is unstable against local perturbations that break
its symmetry, and it decays into a configuration with three
domains separated by two domain walls, as shown in 
Fig.~\ref{asj}. Two domains with the same vacuum 
expectation value connect, and the resulting two domain
walls are pushed out of the center.

\begin{figure}[!h]
\begin{center}
\includegraphics[width=7cm]{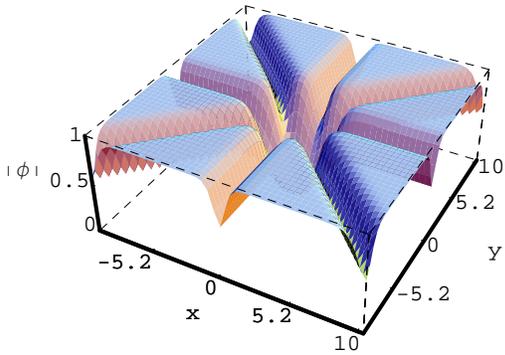}
\end{center}
\caption{Modulus of $\phi$ for the non-BPS junction with winding number two.}
\label{s8j}
\end{figure}

The second type of non-BPS walls we discuss is the eight-junction
shown schematically in Fig.~\ref{N4notBPS}~b). Compared to the
BPS four-junction, this junction has winding number two. We
again found by numerical simulation of the second order
equations of motion that a static configuration of this
type exists.
We show the modulus of the field $\phi$ for this configuration
in Fig.~\ref{s8j}.

\begin{figure}[!h]
\begin{center}
\includegraphics[width=7cm]{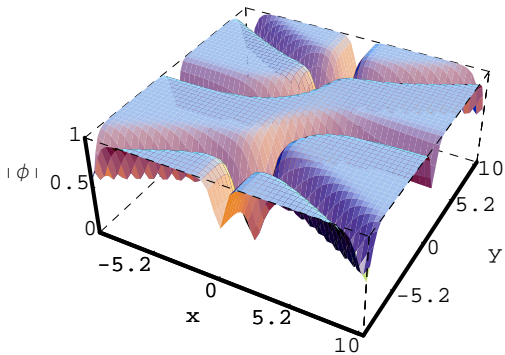}
\end{center}
\caption{Modulus of $\phi$ after the non-BPS eight-junction has decayed into a
configuration with two type~A junctions.}
\label{a8j}
\end{figure}

We also observed that this junction can decay
into two four-junctions with winding number one when a
symmetry breaking perturbation is introduced. The two
resulting junctions repel each other, as shown in
Fig.~\ref{a8j}
We noticed that the eight-junction only decays when the
perturbation is sufficiently large. For now, we leave
open the question whether this indicates that the junction
is meta-stable, or that this behavior is an artifact of
the finite size of  the lattice we used.

\subsection{Numerical methods.}\label{sec:nm}
We simulated the second order equations of motion on a lattice 
using a forward predicting algorithm. The lattice spacing was
chosen to be much smaller than the size of the junctions, and, at
the same time, the size of the lattice was much larger than the
size of the junctions. For our purposes, a lattice of $251$
by $251$ points offered sufficient resolution. 

We added a damping
term to the equations of motion which allows the field to
relax to a minimum energy configuration. At the same time, this
damping stabilizes the forward predicting scheme. 

In order
to generate unstable junctions, we started our simulation with
an initial configuration that was invariant under the same
symmetry transformations as the boundary condition. As these
symmetry are not broken by the equations of motion, the final
configuration is the lowest energy configuration consistent
with the symmetries. Such a configuration may be unstable against
local perturbations which break the symmetry. 

\section{Summary}\label{sec:s}

In this paper we have studied in detail domain wall junctions
in a generalized Wess-Zumino model. We have presented a method
to identify all potentially BPS saturated junctions, and we
have described a procedure to determine whether these junctions
indeed satisfy the BPS bound.
We showed that in the case $N=4$ (the lowest value of $N$
for which there is more than one potential BPS junctions), these 
junctions are in fact BPS saturated.

On the basis of our results, in conjunction with the results
for the basic junction for generic value of $N$ \cite{SV} and
the large $N$ results for the basic junction \cite{GS}, we
speculate that {\it every} potential BPS saturated junction for any $N$ in
this model indeed saturates the BPS bound.    

%\section*{Note added}

\acknowledgements
{Useful discussion with M. Shifman and V. Vento are
gratefully acknowledged. One of us (D.B.) would like to acknowledge warm
hospitality extended to him at the Theoretical Physics Institute of the
University of Minnesota, where part of this work was realized. 
This work was supported in part by the Department 
of Energy under Grant No. de-FG-94ER40823, and by Ministerio de Educaci\'on y
Cultura under
Grant No. DGICYT-PB97-1227.}

\end{document}